# Mapping mechanism between density of states and ultraviolet-visible light absorption spectra

Yang Ling[1], Zhengxin Chen[2], Jiang Wu[1,2,*]

1. School of Energy and Power Engineering, University of Shanghai for Science and Technology, Shanghai 200093, China.

2. College of Energy and Mechanical Engineering, Shanghai University of Electric Power, Shanghai 200090, China.

*Corresponding authors: Jiang Wu (Ph.D., Professor)

E-mail: wjcfd2002@163.com

Postal Address: No. 2103 Pingliang Road, Shanghai 200090, China.

**Abstract:** This paper constructs the mapping relation from density of states (DOS) to UV-vis spectra by using an *ab initio* perspective. Taking $BiOIO_3$ semiconductor photocatalyst as an example, the experimental verification was also carried out. The optical response of the material considers the superposition benefits of all possible transitions. Directly using the difference between valence band maximum (VBM) and conduction band minimum (CBM) will lead to an underestimate of the energy band gap. This paper provides a new idea of linking the density functional theory (DFT) data with experiment phenomenon.

**1 Introduction**

Optical properties and band structure are important physical properties of semiconductor photocatalysts. Experimentally, such as the absorption characteristics under a certain wavelength of light, constitute the UV-vis spectra [1]. Or a photoluminescence (PL) produced by light radiation at a given excitation wavelength [2]. Theoretically, the density functional theory (DFT) calculation can give the density of states (DOS) of semiconductor, and then calculate the energy band gap [3]. While, building a bridge between theory and experiment has always been the interest of researchers.

Inspired by the extended Hückel theory (EHT) to explain the optical properties of small clusters [4], the focus of this paper is to try constructing the mapping relationship from DOS to UV-vis spectra. Besides, a common bismuth-based photocatalyst $BiOIO_3$ [5] was prepared as an example for experimental verification.

**2 Method**s

According to the basic assumption of quantum mechanics, the macroscopic physical properties of a given system can be determined when the specific analytical expression of the wave function is derived. The wave function $\Psi(\vec{r},t)$ of the system satisfies the Schrödinger equation (Eq. 1) under a given potential field $V(\vec{r},t)$.

$$i\hbar \frac{\partial \Psi(\vec{r},t)}{\partial t} = -\frac{\hbar^2}{2m} \nabla^2 \Psi(\vec{r},t) + V(\vec{r},t) \Psi(\vec{r},t) \tag{1}$$

where $\hbar$ is the reduced Planck constant ($\hbar = \frac{h}{2\pi}$), $m$ is the mass of the particle, $i$ is the imaginary unit, and $\nabla^2$ is the Laplace operator. When the potential function of the system does not change with time ($V = V(\vec{r})$), it can be concluded that the stationary wave function $\psi(\vec{r})$ should satisfy the time-independent Schrödinger equation by separating variables method (Eq. 2).

$$E\psi(\vec{r}) = -\frac{\hbar^2}{2m} \nabla^2 \psi(\vec{r}) + V(\vec{r}) \psi(\vec{r}) \tag{2}$$

where $E$ is the total energy of the system. It is difficult to solve the Schrödinger equation directly as the number of electrons increases rapidly. While, by using the extended Hückel theory (EHT) [6], the calculation can be greatly reduced by ignoring the core electron integral that is not involved in chemical reactions, and considering only the valence electrons. When the analytic expressions of wave functions for each energy level are obtained, the transition caused by light radiation from energy level $\alpha$ to $\beta$ can be calculated by oscillator strength (Eq. 3) [4,7].

$$f_{\alpha \to \beta} = \frac{2}{3}\frac{m_e}{\hbar^2}(E_\beta - E_\alpha) \sum_{\xi=x,y,z} \left| \langle \psi(\vec{r}) | \xi \psi(\vec{r}) \rangle \right|^2 \quad (3)$$

where, $m_e$ is the mass of the electron. By superimposing all possible level transitions, the absorption response curve of the system to a certain wavelength of light can be obtained (Eq. 4).

$$A(\lambda) \propto \sum_{\alpha,\beta} f_{\alpha \to \beta} \Gamma_{\alpha \to \beta}(\lambda) \quad (4)$$

where, $\Gamma_{\alpha \to \beta}(\lambda)$ is transition lineshape. The summation takes all possible transitions, where $\alpha$ is taken form the occupied electron state and $\beta$ is taken from the empty orbital.

Different from the wave function method, the density function theory (DFT) calculation cleverly replaces the multi-electron wave function with the electron density in real space, thus greatly reducing the computational dimensions. DFT software usually gives the density of states (DOS) of the system, which reflects the number of quantum states per unit energy level, and is widely used to analyze the optical properties, chemical bond bonding, charge transfer, etc. Combined with the optical response formula given by Eq. 3 and 4, one would be inspired to map directly to UV-vis spectra through DOS data, as shown in Fig. 1. By *ab initio* calculation, the energy level distribution of the system is obtained. When the total number of energy levels is excessive enough, the energy level lines will become too dense to distinguish. In this case, the energy level distribution is represented by the density of states (DOS), whose abscissa axis is the energy distribution (eV), and the ordinate axis is the number of electronic states per unit energy (electrons/eV). In order to apply the procedure of calculating optical properties in extended Hückel theory, we can use the idea of infinitesimal method of calculus to divide and discretize the DOS graph. A discretized micro unit (rectangle) is taken as the object for analysis, while its width is the discrete energy interval $dE$ (0.1~2 eV in this paper for convenient discussion), and its height is the local electron number per unit energy $n(E)$. Then, the area of the rectangle $n(E)dE$ is exactly the number of electrons or vacant orbitals contained in the micro unit. In addition, in the DFT calculation, the orbital of each electron in the system is not given, but only have the charge density of the whole real space, so we assume that the wave function of each electron is orthogonal normalization, and ignore the latter part of the formula, and obtain Eq. 5.

$$A(\lambda) \propto \sum_{\alpha,\beta} [n_\alpha(E)dE][n_\beta(E)dE] \frac{2}{3} \frac{m_e}{\hbar^2} (E_\beta - E_\alpha) \Gamma_{\alpha \to \beta}(\lambda) \tag{5}$$

where, the correction coefficient before the oscillator strength reflects the number of electronic states in the small rectangular cell. The lineshape function is selected as the Gaussian distribution, also known as normal distribution, to expand the oscillator strength (Eq. 6).

$$\Gamma_{\alpha \to \beta}(\lambda) = \frac{1}{\sqrt{2\pi}\sigma} e^{-\frac{(\lambda - \lambda_{\alpha \to \beta})^2}{2\sigma^2}} \tag{6}$$

where, $\sigma$ is the standard deviation. Here, the value is set to 20, which achieves good prediction effect. $\lambda_{\alpha \to \beta}$ is the wavelength at which the current transition occurs, and when the energy unit is eV and the wavelength unit is nm, $\lambda_{\alpha \to \beta}$ is approximately equal to $\frac{1240}{E_\beta - E_\alpha}$.

The raw data of the density of states for a given crystal was computed based on density functional theory via CASTEP code [8]. The Perdew-Burke-Ernzerhof (PBE) form under the generalized gradient approximation (GGA) functional was adopted [9]. And the convergence tolerance is set as: the maximum energy change of 5.0*10$^{-5}$ eV/atom, the maximum force of 0.1 eV/Å, the maximum stress of 0.2 GPa, and the maximum displacement 0.005 Å.

In order to combine the theory with experiment, the BiOIO$_3$ semiconductor photocatalyst was taken as an example to carry out the experimental study, which is a common bismuth-based photocatalyst that widely used in carbon dioxide reduction and heavy metal removal [10,11]. The preparation method was referred to previous work [12], and please see details in the *Supplementary Material*. The UV-vis spectra of the prepared $BiOIO_3$ crystal was obtained by an ultraviolet-visible spectrophotometer (Shimadzu UV-3600 Plus).

**3 Results and discussions**

As shown in Fig. 2a is the schematic diagram of the crystal cell of $BiOIO_3$, which presents a typical Aurivillius-type layered structure [5], with $(Bi_2O_2)^{2+}$ layers as the skeleton and $(IO_3)^-$ layers suspended on both sides, and layers interact with each other by van der Waals forces. Fig. 2b shows the density of states (DOS) of $BiOIO_3$ crystal calculated via DFT, and the band gap is only 1.685 eV, which is significantly underestimated compared with the experimental value (about 3 eV, [10-12]). Researchers often attribute this to the incomplete characterization of semiconductor band gap by the PBE-GGA functional used in DFT calculation [13]. Empirical parameter correction seems to deviate from the original intention of the DFT *ab initio* ideology, since it essentially hopes to predict the macroscopic properties of the system from a few theoretical physical parameters. However, let's recall that the band gap given by traditional DFT software is generally the difference between valence band maximum (VBM) and conduction band minimum (CBM). While, after a careful observation of Eq. 4, the optical absorption characteristics of the system, it can be found that it considers the comprehensive effect formed by the superposition of all possible energy level transitions. Therefore, the energy level transitions of VBM and CBM alone contribute very little to the overall UV-vis absorption spectra.

Fig. 2c is the theoretically predicted UV-vis spectra of $BiOIO_3$ by using Eq. 5, based on the DOS data calculated via DFT. It can be seen that the interval $dE$ selected during DOS discretization has a great influence on the structure of UV-vis spectra images. When $dE$ is relatively large (2 eV), the image is loose and the detail description is not perfect due to the small number of sample points selected. While, with the gradual decrease of $dE$ (2→0.1 eV), the UV-vis image becomes more and more delicate, and the characteristic peak is clearly recognizable. It is predicted that $BiOIO_3$ has typical semiconductor characteristics and strong absorption of ultraviolet light.

Fig. 3a shows the XRD pattern of the $BiOIO_3$ crystal, which sharp and obvious characteristic peak indicates the successful synthesis of the sample (ICSD#262019) [5,10]. $BiOIO_3$ prepared by hydrothermal method presents the microscopic morphology of nanosheets (Fig. 3b). As can be seen from the UV-vis spectroscopic experimental results (Fig. 3c), $BiOIO_3$ behaves typically as a semiconductor, responding only to certain wavelengths of light. The absorption of light is mainly distributed in the ultraviolet region. The absorption curve of light decreases at about 300 nm, and remains at very low levels after 400 nm. There is almost no response to light in the far infrared region. The experimental results are very similar to those predicted by theory, which confirms the feasibility of the proposed mapping method reported in this paper.

**4 Conclusions**

In summary, the UV-vis spectra of $BiOIO_3$ crystal was successfully predicted by discretizing the DOS graph into transition element unit, which was inspired by the oscillator strength. Estimation of the band gap width using the difference between VBM and CBM is understated, because the optical response examines the superposition of all possible transitions. This paper provides a new idea for DFT calculation to better relate experimental phenomena.

**Acknowledgments**

This work was partially sponsored by National Natural Science Foundation of China (52076126)

**References**

[1] P. Makuła, M. Pacia, W. Macyk, J. Phys. Chem. Lett. 9 (2018) 6814-6817.

[2] T. Jia, J. Wu, Z. Ji, C. Peng, Q. Liu, M. Shi, J. Zhu, H. Wang, D. Liu, M. Zhou, Appl. Catal. B-Environ. 284 (2021) 119727.

[3] D.J. Martin, K. Qiu, S.A. Shevlin, A.D. Handoko, X. Chen, Z. Guo, J. Tang, Angew. Chem. Int. 53 (2014) 9240-9245.

[4] N.V. Tepliakov, E.V. Kundelev, P.D. Khavlyuk, Y. Xiong, M.Y. Leonov, W. Zhu, A.V. Baranov, A.V. Fedorov, A.L. Rogach, I.D. Rukhlenko, ACS Nano, 13 (2019) 10737-10744.

[5] S.D. Nguyen, J. Yeon, S. Kim, P.S. Halasyamani, J. Am. Chem. Soc. 133 (2011) 12422-12425.

[6] R. Hoffmann, J. Chem. Phys. 39 (1963) 1397-1412.

[7] Z.K. Tang, A. Yanase, T. Yasui, Y. Segawa, Phys. Rev. Lett. 71 (1993) 1431-1434.

[8] S.J. Clark, M.D. Segall, C.J. Pickard, P.J. Hasnip, M.I.J. Probert, K. Refson, M.C. Payne, Z. Krist-Cryst. Mater. 220 (2005) 567-570.

[9] J. Perdew, K. Burke, M. Ernzerhof, Phys. Rev. Lett. 77 (1996) 3865-3868.

[10] F. Chen, H. Huang, L. Ye, T. Zhang, Y. Zhang, X. Han, T. Ma, Adv. Funct. Mater. 28 (2018) 1804284.

[11] X. Sun, J. Wu, Q. Li, Q. Liu, Y. Qi, L. You, Z. Ji, P. He, P. Sheng, J. Ren, W. Zhang, J. Lu, J. Zhang, Appl. Catal. B-Environ. 218 (2017) 80-90.

[12] X.M. Qi, M.L. Gu, X.Y. Zhu, J. Wu, H.M. Long, K. He, Q. Wu, Chem. Eng. J. 285 (2016) 11-19.

[13] A.J. Garza, G.E. Scuseria, J. Phys. Chem. Lett. 7 (2016) 4165-4170.

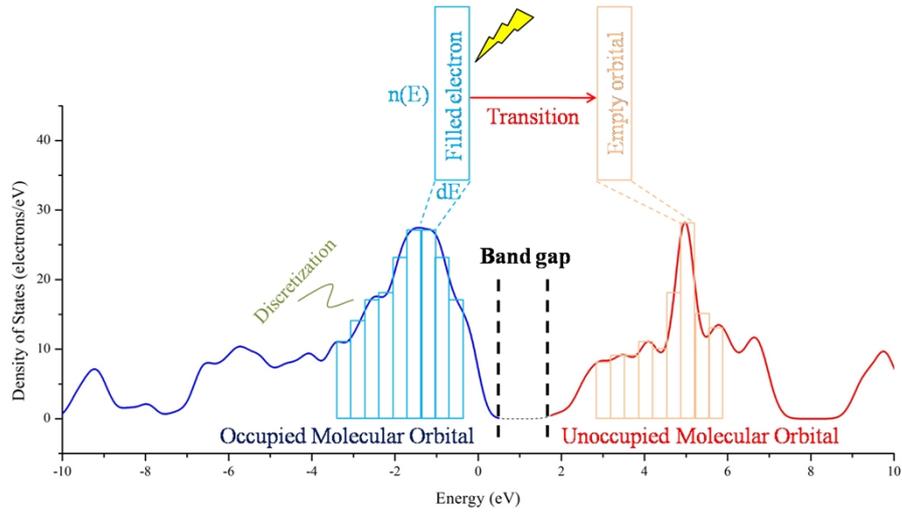

**Fig. 1.** Schematic diagram of mapping from density of states to UV-vis spectra.

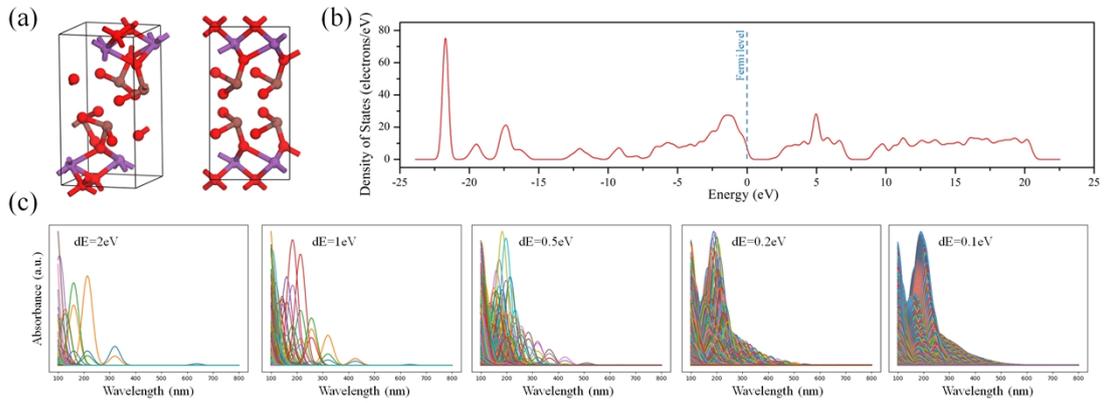

**Fig. 2.** (a) Crystal structure of $BiOIO_3$; (b) Density of states of $BiOIO_3$ achieved by DFT calculation; (c) Theoretical prediction of the UV-vis spectra.

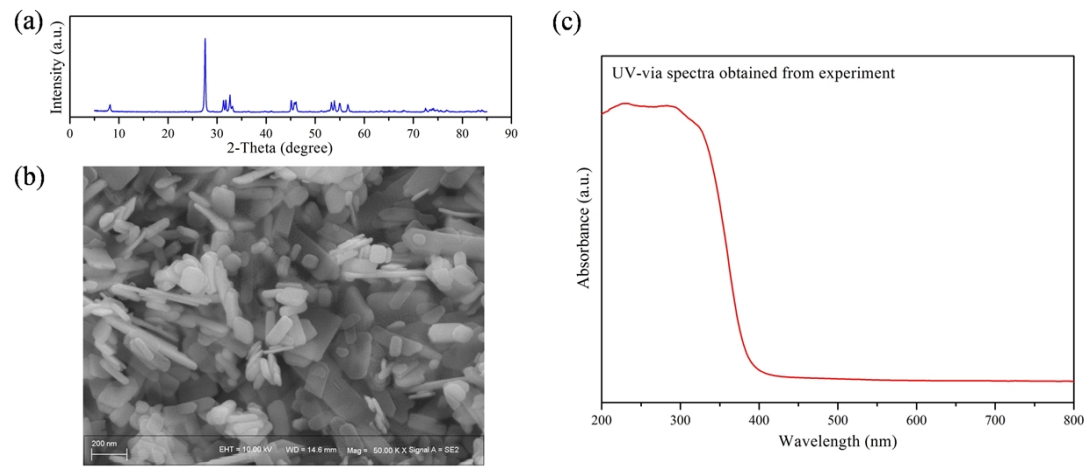

**Fig. 3.** Experimental data of $BiOIO_3$: (a) XRD pattern; (b) SEM image; (c) UV-vis spectra.